\documentclass[10pt,twocolumn]{article}

\usepackage[letterpaper,margin=0.75in,columnsep=0.25in]{geometry}
\usepackage{cite}
\usepackage{amsmath,amssymb,amsfonts}
\usepackage{graphicx}
\usepackage{textcomp}
\usepackage{xcolor}
\usepackage{booktabs}
\usepackage{array}
\usepackage{siunitx}
\usepackage{caption}
\begin{document}

\title{Local Suppression of Superconductivity in Niobium Films by Focused Silicon Ion Implantation}

\author{\parbox{0.92\textwidth}{\centering
Rochelle~Y.~Qu$^{1}$,
Shunsuke~Nishimura$^{1}$,
Max~Pr\"opper$^{1}$,
Joseph~J.~Barrera$^{1}$,
Alexander~Brooks$^{1}$,
Joshua~Vande~Berg$^{1}$,
Achim~Nadzeyka$^{2}$,
Aravind~R.~Sreeja$^{1}$,
Jay~C.~LeFebvre$^{1}$,
Roger~K.~Lake$^{1}$,
and~Shane~A.~Cybart$^{1}$\\[4pt]
\small $^{1}$Department of Electrical and Computer Engineering, University of California, Riverside, CA 92521, USA\\
\small $^{2}$Raith GmbH, Dortmund, Germany\\
\small Corresponding author: \mbox{cybart@ucr.edu}}
}
\date{}

\twocolumn[
\begin{@twocolumnfalse}
\maketitle
\begin{abstract}
Magnetic flux trapped in a superconducting circuit shifts the critical currents of nearby Josephson junctions and disturbs the operation of the circuit. The usual remedy is a moat etched through the ground plane to hold the flux away from the logic. We report direct write silicon ion implantation with a focused ion beam as a way to make moats without etching the film, by locally suppressing superconductivity in niobium (Nb). Single implanted lines were written across Nb electrodes in films of three thicknesses over two orders of magnitude in ion fluence, and the transition temperature, resistance, and critical current density of each line were measured. The transition temperature of the line falls with fluence while the surrounding Nb is unchanged, and the critical current density falls by about two orders of magnitude before the line becomes resistive at 4.2~K. The product of the critical current and the resistance of the lines follows a single power law in reduced temperature. This gives an estimate of the critical current of a line at any temperature from its transition temperature and resistance. A time dependent Ginzburg-Landau simulation indicates that an implanted region can capture and hold a vortex. From these results we suggest a starting ion energy and fluence for implanted moats in a 100~nm Nb ground plane at 4.2~K and estimate the equivalent broad beam fluence.

\end{abstract}

\vspace{1em}
\end{@twocolumnfalse}
]

\section{Introduction}

Data centers in the United States consumed 4.4\% of the nation's electricity in 2023, and this share is projected to reach 6.7\% to 12\% by 2028~\cite{Shehabi2024}. Superconducting digital electronics offers a path to large energy savings. Logic based on Josephson junctions switches with energies that are orders of magnitude lower than those of complementary metal-oxide-semiconductor (CMOS) transistors, and the advantage can remain after the cost of cooling to 4~K is included~\cite{Holmes2013,Tolpygo2016}. Among the engineering challenges that remain, two of the most important are scaling the fabrication process to very large scale integration and mitigating magnetic flux trapping in the circuits~\cite{Tolpygo2016}.

Flux trapping occurs when a superconducting film is cooled through its transition temperature $T_C$ in a residual magnetic field. The flux is trapped as quantized vortices. A vortex near a Josephson junction couples flux into the junction and shifts its critical current ($I_C$)~\cite{Scalapino1967,golod2015}. The standard remedy is the flux trapping moat, a hole or slit through the superconducting ground plane~\cite{Jeffery1995,Semenov2016}. A vortex in a moat costs no superconducting condensation energy. The flux therefore settles in the moat instead of in the film near the logic. Moats are patterned by photolithography and reactive ion etching that removes the superconducting film~\cite{Kapur2026Mitigation,Kapur2026Flux}. They have been integrated into standard cell libraries of superconducting digital logic~\cite{Yamae2025}.

Ion irradiation has a long history as a tool for tuning superconductors. In 1988, ion irradiation was shown to reduce the critical current density ($J_C$) of YBa$_2$Cu$_3$O$_{7-\delta}$ (YBCO) films by three orders of magnitude while $T_C$ remained above 77~K~\cite{White1988}. Since then, focused ion beams have been used to induce disorder, implant ions, and mill both low-$T_C$ and high-$T_C$ superconductors with a variety of ion species~\cite{cybart2015,Cybart2014,Ogawa2021,Brooks2025,Hoflich2023}.

Ion implantation offers an alternative to etched moats that follows the path taken by the CMOS industry, where implantation is the standard way to modify a material locally during fabrication~\cite{Williams1998}. The key to this approach is that the implanted ions, and the lattice displacements they create, suppress superconductivity locally. The implanted region has a reduced critical current density, or becomes normal, at the operating temperature while the surrounding film is unchanged. A vortex lowers its energy at such a site, so the implanted region pins flux in the same manner as a moat. An implanted moat does not require etching a hole through the film. This may simplify the planarization that the most advanced superconducting processes apply to every layer~\cite{Tolpygo2015}.

In this paper, we use a focused beam of silicon (Si) ions to measure how the implanted fluence changes the transport properties of niobium (Nb). The focused beam writes each fluence directly, without a mask, so many fluences can be compared on a single chip in a single cooldown. From these results, we suggest the ion energy and fluence for implanted moats in 4~K logic.

\section{Experimental Methods}

Two implantation experiments were performed with a focused beam of silicon ions. We chose silicon because it is compatible with Nb, whereas the gallium used in most focused ion beam (FIB) systems is not~\cite{Datesman2005}. Experiment 1 used a low beam current in a single pass on a 200~nm film. Experiment 2 used a higher beam current in repeated scans on 50, 100, and 200~nm films.

\subsection{Samples}

\begin{figure}[!tbp]
 \centering
 \includegraphics[width=0.75\columnwidth]{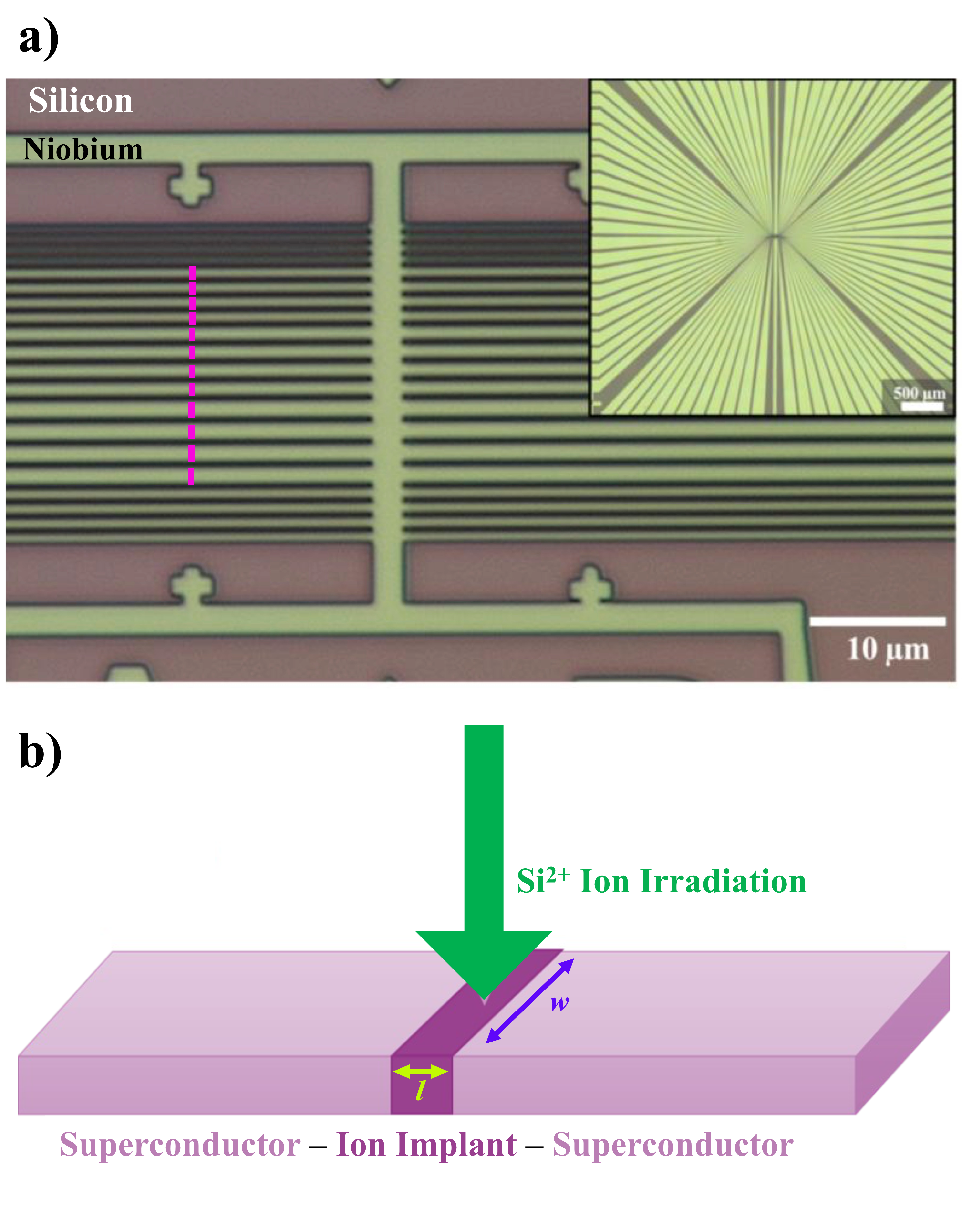}
 \caption{Sample layout and implant geometry. (a) Optical micrograph of the center of a niobium chip. The electrodes appear yellow and the exposed substrate gray. The pink marks show where the Si lines were implanted, and the inset shows the full 5~mm by 5~mm chip. (b) Schematic of an electrode of width $w$ with an implanted line of lateral extent $l$ written across its full width. Not to scale.}
 \label{fig:Sample}
\end{figure}

The samples were \qtyproduct{5x5}{\milli\metre} chips cut from Nb films deposited on silicon substrates with a SiO$_2$ dielectric layer and patterned by MIT Lincoln Laboratory in a single layer Nb process. Both experiments used the same circuit design, with 40 four terminal electrodes per chip, 250 to 1000~nm wide. The 20 electrodes on the left half of the chip are labeled A and those on the right half B. Figure~\ref{fig:Sample}(a) shows the electrodes at the center of the chip, and the inset shows the full chip. Figure~\ref{fig:Sample}(b) shows the geometry of an implanted line, which crosses the full width $w$ of the electrode and has a lateral extent $l$ along it. Experiment 1 used a 200~nm film, and its electrical measurements were made on the 750 and 1000~nm electrodes. Experiment 2 used 50, 100, and 200~nm films and electrodes of all four widths. An unimplanted reference electrode on the 200~nm film, \SI{2}{\micro\meter} wide and \SI{30}{\micro\meter} long, was measured for comparison.

\subsection{Focused Ion Beam Implantation}

Silicon implantation was performed with a Raith VELION FIB system. The VELION uses a liquid metal alloy ion source that emits several ion species and charge states from a single eutectic source. A Wien filter selects the species and charge state by mass-to-charge ratio for injection into the column~\cite{Bischoff2005,Bischoff2016}. We chose doubly ionized silicon, Si$^{2+}$, because it reaches twice the energy of Si$^{+}$ at the same column voltage. At 35~kV, the incident ion energy was 70~keV. Direct write patterning allowed the fluence to be varied between electrodes on the same chip without additional lithography. The VELION specifies the dose of a line scan as a line fluence, the number of implanted Si ions per unit length of the line (ions/cm), and we quote line fluence throughout this article.

Table~\ref{tab:write} lists the write parameters of the two experiments. In Experiment 1, the beam delivered the full fluence at each point before stepping to the next. In Experiment 2, each line was scanned repeatedly in the same direction, with approximately 300 ions delivered per visit. The write modes therefore differ primarily in local dose rate.
\begin{table}[!b]
\caption{Write Parameters of the Two Implantation Experiments}
\label{tab:write}
\centering
\footnotesize
\begin{tabular}{@{}>{\raggedright\arraybackslash}p{2.3cm}p{2.9cm}p{2.5cm}@{}}
\toprule
 & Experiment 1 & Experiment 2 \\
\midrule
Ion and energy & Si$^{2+}$, 70~keV & Si$^{2+}$, 70~keV \\
Beam current & 1.8 pA & 22 pA \\
Beam diameter & $\approx$6~nm & $\approx$10 to 12~nm \\
Aperture & \SI{10}{\micro\meter} & \SI{10}{\micro\meter} \\
Working distance & 7.7 mm & 7.7 mm \\
Field of view & \SI{200}{\micro\meter} & \SI{200}{\micro\meter} \\
Point spacing & 2~nm & 3~nm \\
Scan & single pass, full fluence at each point & repeated scans, same direction \\
Dwell per point & 18 ms to 0.36 s$^{\ast}$ & \SI{4.4}{\micro\second} \\
Ions per visit & $1\times10^{5}$ to $2\times10^{6}$$^{\ast}$ & $\approx$300$^{\ast}$ \\
Line fluence (ions/cm) & $5\times10^{11}$ to $1\times10^{13}$ & $1\times10^{11}$ to $5\times10^{12}$ \\
Area fluence (ions/cm$^{2}$)$^{\dagger}$ & $8.3\times10^{17}$ to $1.7\times10^{19}$ & $9.1\times10^{16}$ to $4.5\times10^{18}$ \\
Nb thickness & 200~nm & 50, 100, 200~nm \\
Electrode width & 750, 1000~nm$^{\ddagger}$ & 250 to 1000~nm \\
Write time & $\approx$9 s to 3 min per \SI{1}{\micro\meter} line$^{\ast}$ & $\approx$90 s for all 40 lines \\
\bottomrule
\multicolumn{3}{@{}p{7.7cm}@{}}{$^{\ast}$Calculated from the beam current and point spacing. Si$^{2+}$ at 1.8 pA delivers $5.6\times10^{6}$ ions/s, and at 22 pA, $6.9\times10^{7}$ ions/s. $^{\dagger}$Line fluence divided by the beam diameter, 6~nm and 11~nm. $^{\ddagger}$Electrical measurements; the AFM used 500~nm electrodes on both chips.}
\end{tabular}
\end{table}

Because the beam diameters differ, the same line fluence produces different area fluences. We therefore also report the area fluence in the beam spot, which is the line fluence divided by the beam diameter, 6~nm in Experiment 1 and 11~nm in Experiment 2. At equal line fluence, Experiment 1 delivers approximately twice the area fluence.

\subsection{Measurements}\label{sec:meas}

The surface topography of the lines on both chips was measured after the electrical measurements with tapping mode atomic force microscopy (AFM) using a probe with a nominal tip radius of 8~nm. We define the trench depth of a line as the difference between the height of the electrode on either side of the line and the minimum of the height profile, with the profile averaged along the line over the width of the electrode. Ion ranges and sputtering were simulated with the Stopping and Range of Ions in Matter (SRIM) Monte Carlo code~\cite{Ziegler2010} for 70~keV Si ions incident on a 200~nm Nb film on a SiO$_2$ and Si substrate, for a sweep of the Si energy, and for 100~keV N ions in a 100~nm Nb film for comparison with the literature. Vortex trapping was simulated with the two dimensional time dependent Ginzburg-Landau equations described in Section~\ref{sec:tdgl}.

Resistance versus temperature, $R(T)$, was measured between 2 and 300~K in a four terminal configuration in a cryostat with a 9~T superconducting magnet, using a \SI{1}{\micro\ampere} root mean square excitation at 17.3~Hz supplied through a \SI{1}{\mega\ohm} series resistor and two lock-in amplifiers for the electrode voltage and the excitation current. We define the transition temperature $T_C$ of an implanted line as the temperature of the peak in $dR/dT$ of the line transition, and we quote the full width at half maximum of that peak as the width of the transition. The $R(T)$ data were interpolated onto a 0.01~K grid, smoothed with a Gaussian kernel of standard deviation 0.04~K, and differentiated numerically. Where no $R(T)$ was measured, we bracket $T_C$ between the highest temperature at which the line carries a supercurrent and the lowest temperature at which it is resistive at low bias. For the comparison of Section~\ref{sec:moats}, $T_C$ of an Experiment 2 line is the midpoint of this bracket, and $T_C$ of an Experiment 1 line is the temperature at which a linear fit of $J_C(T)$ extrapolates to zero.

The critical current $I_C$ was determined from the current-voltage ($I$-$V$) characteristics as the current at which the voltage first exceeded \SI{10}{\micro\volt}, or five times the standard deviation of the voltage noise if that is larger, on the branch of the sweep where the magnitude of the current increases. The values for the two polarities were averaged. The critical current density is $J_C = I_C/A$, where $A$ is the nominal cross section of the electrode, its width times the film thickness. We use the nominal cross section throughout because the trench depth was measured only on the 200~nm chips. The resistance $R$ of a line is the slope of its $I$-$V$ characteristic over the highest 15\% of the bias range in Experiment 1, and its low bias resistance at 8.5~K, just below the film transition, in Experiment 2. We restricted the quantitative analysis of Experiment 2 to electrodes 500~nm wide or wider, because the 250~nm electrodes behaved inconsistently on every chip. This is not understood and merits additional study.

\section{Results and Discussion}

\subsection{Surface Topography}\label{sec:3A}

\begin{figure}[!tbp]
 \centering
 \includegraphics[width=.8\columnwidth]{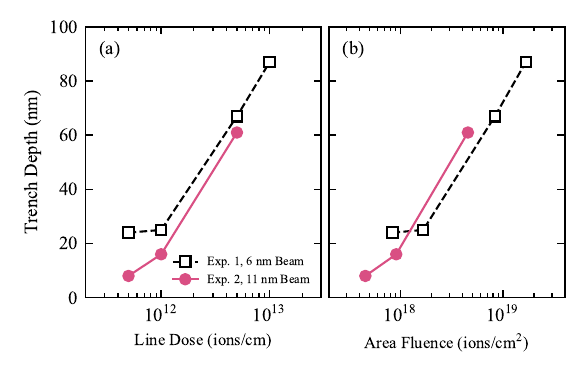}
 \caption{Depth of the trench sputtered by the focused beam on 500~nm wide electrodes, measured by atomic force microscopy (AFM), as a function of (a) the line fluence and (b) the area fluence in the beam spot, which is the line fluence divided by the beam diameter. Open squares are Experiment 1 with the 6~nm beam, and filled circles are Experiment 2 with the 11~nm beam, both on 200~nm films. The depths are approximate and were read from height profiles averaged over the electrode width.}
 \label{fig:trench}
\end{figure}

Figure~\ref{fig:trench} shows the trench depth on the 500~nm wide B side electrodes of the 200~nm chips of both experiments. In Experiment 1, the depth increases from approximately 24~nm at $5\times10^{11}$~ions/cm to 87~nm at $1\times10^{13}$~ions/cm. In Experiment 2, it increases from approximately 8 to 61~nm between $5\times10^{11}$ and $5\times10^{12}$~ions/cm. The Experiment 2 trenches are shallower at equal line fluence [Fig.~\ref{fig:trench}(a)]. When plotted against the area fluence in the beam spot [Fig.~\ref{fig:trench}(b)], the two experiments fall on one curve. This is consistent with sputtering that depends on the number of ions per unit area, because the approximately twice wider Experiment 2 beam spreads the ions over twice the area. Finally, the two lowest Experiment 1 fluences yield similar depths. This may possibly be due to a surface contamination floor observed on this chip.

\begin{figure}[!tbp]
 \centering
 \includegraphics[width=.7\columnwidth]{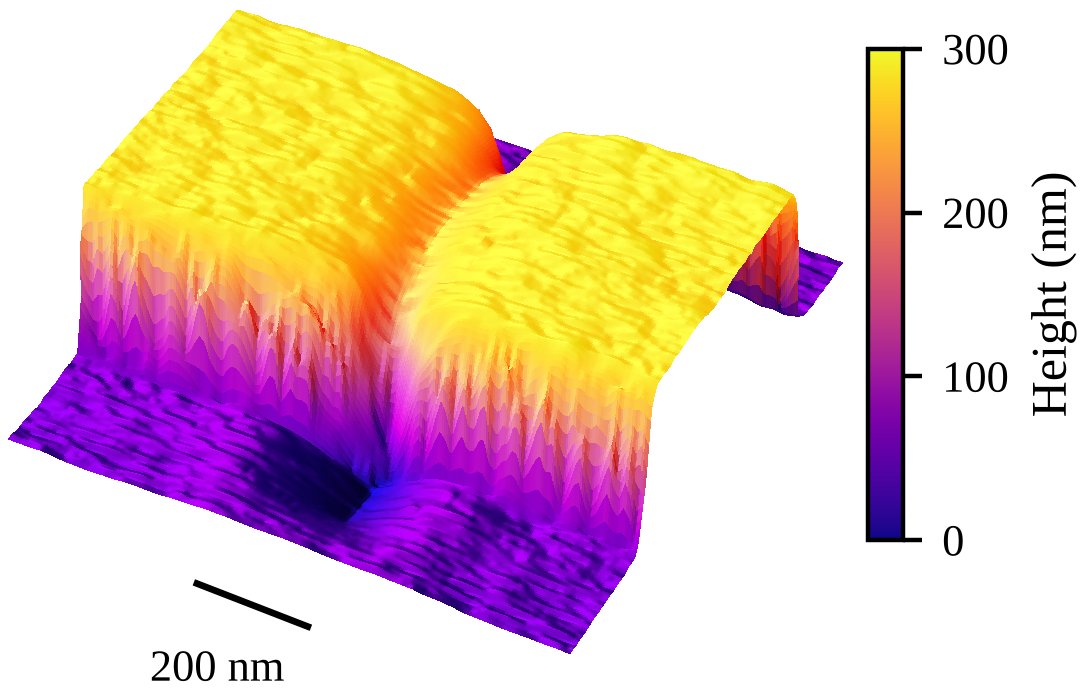}
\caption{Atomic force microscopy (AFM) topography of a trench written in Experiment 2 at $5\times10^{12}$~ions/cm across a 500~nm wide electrode in a 200~nm film, shown over the full $1\times1$~\si{\micro\meter} scan. Heights are referenced to the lowest point of the scan. The implanted line continues into the substrate on both sides of the electrode. The trench depth averaged across the electrode is about 61~nm. The surface is interpolated for display, and the vertical scale is exaggerated 1.4-fold.}
 \label{fig:afm}
\end{figure}

Figure~\ref{fig:afm} shows the AFM topography of the deepest Experiment 2 trench, written at $5\times10^{12}$~ions/cm across a 500~nm electrode of the 200~nm film. The implanted line continues into the substrate on both sides of the electrode. The trench depth averaged across the electrode is about 61~nm. The trench width at half depth is 50 to 60~nm in Experiment 1 and 45 to 80~nm in Experiment 2. The bases of the deep trenches broaden to approximately 200~nm, partly because of tip effects. The 8~nm AFM tip cannot fully trace these narrow, deep trenches, so their true widths may be smaller and their depths greater.

\subsection{Ion Range and Sputtering}\label{sec:3b}

SRIM gives a projected range of 45.2~nm for 70~keV Si in Nb, so the ions come to rest within approximately the upper 100~nm of the 200~nm film and none reach the substrate. The lateral straggle sets how wide the implanted line is compared with the 6~nm beam. The region containing the central 90\% of the simulated displacement events is approximately 53~nm wide. This agrees with the measured trench width at half depth of 50 to 60~nm (Section~\ref{sec:3A}), and we take $l = 53$~nm as the lateral extent of the implanted line. Because SRIM assumes a static, flat, amorphous target, these values describe the first ions to arrive.

Dividing the line fluence by this lateral extent gives the area density of implanted ions after straggle, which is the quantity a broad beam sets directly. At $1\times10^{13}$~ions/cm it is approximately $1.9\times10^{18}$~ions/cm$^2$, compared with $1.7\times10^{19}$~ions/cm$^2$ in the beam spot. A SRIM sweep of the ion energy places the projected range of Si in Nb at 100~nm for approximately 162~keV.

SRIM predicts a Nb sputter yield of approximately 1.33 atoms per incident 70~keV Si ion. At the maximum line fluence, this yield predicts a removed cross section of approximately $2.39\times10^{4}$~nm$^2$. The measured trench at this fluence (Fig.~\ref{fig:trench}), approximated as a triangle 200~nm wide and 87~nm deep, has a cross section of approximately $9\times10^{3}$~nm$^2$, within a factor of 3 of the prediction. We attribute the difference to the fall of the sputter yield as the surface becomes silicon rich. In steady state each ion adds one silicon atom and sputters 1.33 atoms, so the net removal of niobium falls to about 0.33 atom per ion.

\subsection{Resistance Versus Temperature}\label{sec:rt}

\begin{figure}[!tbp]
 \centering
 \includegraphics[width=.9\columnwidth]{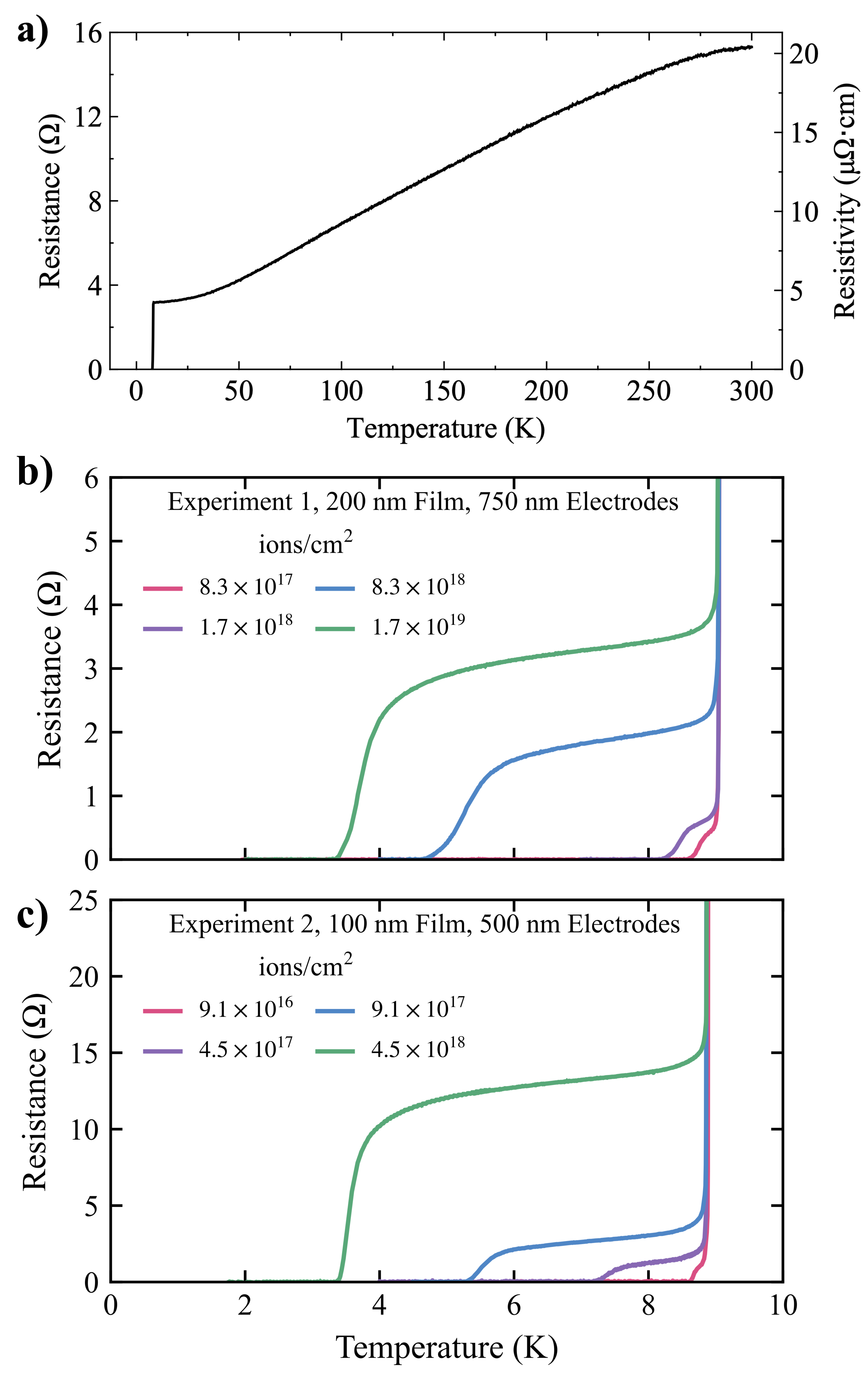}
 \caption{Resistance as a function of temperature. (a) Unimplanted Nb reference electrode on the 200~nm film, with the resistivity on the right axis. The film has a single superconducting transition near 9~K. (b) Experiment 1: 200~nm thick film written in a single pass, 750~nm wide electrodes. (c) Experiment 2: 100~nm thick film written with repeated scans, 500~nm wide electrodes. The line fluences are $5\times10^{11}$, $1\times10^{12}$, $5\times10^{12}$, and $1\times10^{13}$~ions/cm in (b) and $1\times10^{11}$, $5\times10^{11}$, $1\times10^{12}$, and $5\times10^{12}$~ions/cm in (c). Colors give the area fluence in the beam spot in ions/cm$^2$, which is the line fluence divided by the 6~nm and 11~nm beam diameters. A constant offset was subtracted from each curve. The transition near 9~K is that of the unimplanted niobium. The plateau below it is the normal state resistance of the implanted line, and the second transition at lower temperature is that of the line.}
 \label{fig:rt}
\end{figure}

Figure~\ref{fig:rt}(a) shows $R(T)$ of the unimplanted reference electrode on the 200~nm film between 2 and 300~K, with the resistivity $\rho = Rwt/L$ on the right axis, where $w$, $t$, and $L$ are the electrode width, film thickness, and length. The film has a single transition near 9~K. Its resistivity is approximately \SI{4}{\micro\ohm\centi\meter} just above $T_C$ and \SI{20}{\micro\ohm\centi\meter} at 300~K.

Figure~\ref{fig:rt}(b) shows $R(T)$ for the 750~nm wide electrodes of Experiment 1 implanted at $5\times10^{11}$, $1\times10^{12}$, $5\times10^{12}$, and $1\times10^{13}$~ions/cm. The 1000~nm wide electrodes behave in the same way and are omitted. Each curve exhibits the film transition near 9~K, a resistance plateau, and a second transition at lower temperature. The plateau is the normal state resistance of the implanted line in series with the superconducting Nb on either side, and the second transition is that of the line. The line $T_C$ is 8.8~K, 8.5 to 8.7~K, 5.3 to 6.0~K, and 3.7 to 3.9~K at the four fluences, with transition widths of 0.2 to 0.6~K. Each range spans the two electrode widths. At $1\times10^{13}$~ions/cm, both electrode widths are resistive at 4.2~K, with approximately 70 to 80\% of the plateau resistance.

The plateau and the second transition also show that no undamaged Nb remains beneath the line. If an undamaged layer remained, it would become superconducting near 9~K and short the line, and neither feature would appear. Two processes act together to modify the full thickness of the film. The beam sputters a trench approximately 87~nm deep at $1\times10^{13}$~ions/cm (Fig.~\ref{fig:trench}), and the ions that arrive as the surface recedes displace atoms to about 100~nm below the receding surface, so the damage reaches about 190~nm below the original surface. The line is therefore both thinner and disordered. The static SRIM range of Section~\ref{sec:3b} describes only the first ions to arrive. Channeling~\cite{Tzanev1992} and the fluence dependent evolution of the target~\cite{Moller2014} may extend the damage through the remaining thickness.

Figure~\ref{fig:rt}(c) shows $R(T)$ for the 500~nm wide electrodes in the 100~nm film of Experiment 2 at $1\times10^{11}$, $5\times10^{11}$, $1\times10^{12}$, and $5\times10^{12}$~ions/cm. The same two transitions and the plateau between them appear. The line $T_C$ is 7.4~K (width 0.3~K) at $5\times10^{11}$ and 5.5~K (width 0.4~K) at $1\times10^{12}$~ions/cm, and the plateau at $1\times10^{12}$~ions/cm is 3.3~$\Omega$. At $5\times10^{12}$~ions/cm, the line transition lies between 3 and 4~K, in agreement with the $I$-$V$ bracket, and the plateau is an order of magnitude higher than at $1\times10^{12}$~ions/cm. At $1\times10^{11}$~ions/cm, no separate line transition appears, and the resistance reaches zero at 8.6 to 8.7~K within the film transition at 8.9~K. In the 50~nm film, every line that was measured is resistive at 4~K. Its $T_C$ is about 3~K at $1\times10^{11}$ and $5\times10^{11}$~ions/cm and below 2~K at $1\times10^{12}$~ions/cm and above. In the 200~nm film, the line transition stays within 1~K of the film transition up to $5\times10^{12}$~ions/cm.

For Experiment 1, we estimate an effective resistivity of the implanted line from the plateau resistance as $\rho_{\mathrm{eff}} = (R\times w)\,t_l/l$, where $t_l$ is the thickness of the film under the line and $l \approx 53$~nm is its lateral extent. The thickness under the line is the film thickness less the trench depth of Fig.~\ref{fig:trench}: 176, 175, 133, and 113~nm at the four fluences. The full remaining thickness is used because no part of the film shunts the line. The result rises from approximately \SI{100}{\micro\ohm\centi\meter} at $5\times10^{11}$~ions/cm to 160, 410, and \SI{590}{\micro\ohm\centi\meter} at $1\times10^{12}$, $5\times10^{12}$, and $1\times10^{13}$~ions/cm. These values are well above the resistivity of the unimplanted film and are comparable to the values of order $10^{2}$~\si{\micro\ohm\centi\meter} reported for ion irradiated Nb~\cite{Kohopaa2024}.

However, the lines retain $T_C$ between approximately 3.7 and 8.8~K, well above the 0.28~K reported for Si rich amorphous Nb$_x$Si$_{1-x}$~\cite{Bishop1985}. This suggests strong disorder in Nb rather than the formation of a uniform NbSi alloy. A composition measurement would resolve this.

The only condition measured in both experiments is a 200~nm film at $5\times10^{12}$~ions/cm. The single pass lines of Experiment 1 have $T_C$ of 5.3 to 6.0~K, and the repeated scan line of Experiment 2 has $T_C$ of 8.1~K. Because the Experiment 2 beam was approximately twice as wide, the area fluences are $8.3\times10^{18}$ and $4.5\times10^{18}$~ions/cm$^2$. Interpolating the Experiment 1 data to the Experiment 2 area fluence gives $T_C \approx 6.8$~K, so a difference of about 1.3~K remains at equal area fluence. The similar trench depths at equal area fluence (Fig.~\ref{fig:trench}) show that the two modes sputter alike, so the remaining difference is not from material removal. It is not understood and merits additional study.

\subsection{Current-Voltage Characteristics}

\begin{figure}[!tbp]
 \centering
 \includegraphics[width=.9\columnwidth]{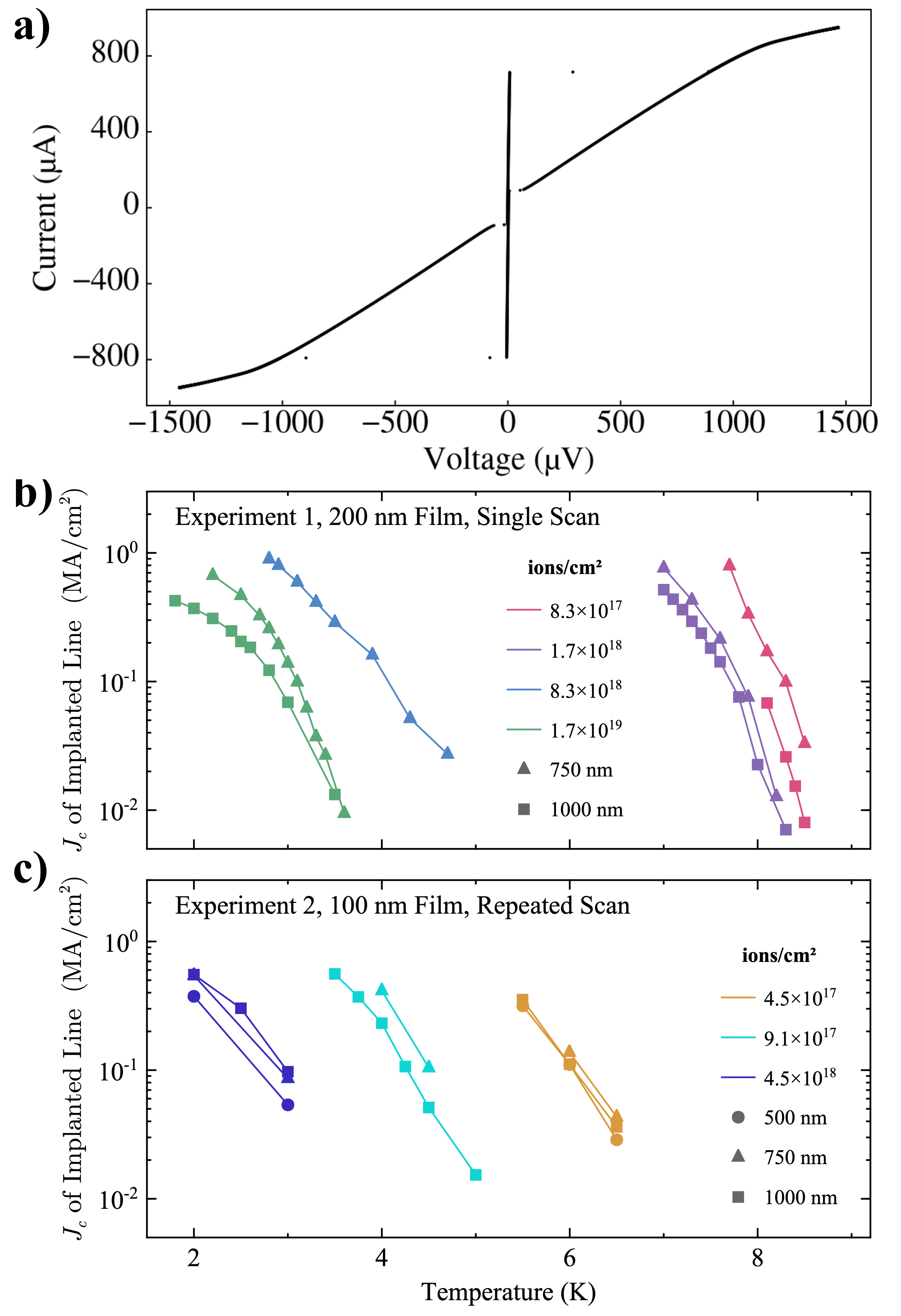}
 \caption{Current-voltage characteristic and critical current density of the implanted lines. (a) $I$-$V$ characteristic at 2~K of a 1000~nm wide niobium electrode with a silicon line implanted at $1\times10^{13}$~ions/cm ($1.7\times10^{19}$~ions/cm$^2$ in the beam spot). The critical current of \SI{750}{\micro\ampere} is the average of the positive and negative switching currents. (b) Critical current density of the implanted lines as a function of temperature in Experiment 1: 200~nm film written in a single pass, for electrodes 750~nm (triangles) and 1000~nm (squares) wide. (c) Experiment 2: 100~nm film written with repeated scans, for electrodes 500~nm (circles), 750~nm (triangles), and 1000~nm (squares) wide. Each curve is one electrode. The line fluences in (c) are $5\times10^{11}$, $1\times10^{12}$, and $5\times10^{12}$~ions/cm. Colors give the area fluence in the beam spot in ions/cm$^2$ with the same scale as Fig.~\ref{fig:rt}.}
 \label{fig:IcT}
\end{figure}

Figure~\ref{fig:IcT}(a) shows the $I$-$V$ characteristic at 2~K of a 1000~nm wide electrode of Experiment 1 implanted at $1\times10^{13}$~ions/cm, below $T_C$ of the line. The positive and negative branches switch at approximately 710 and \SI{790}{\micro\ampere}, and their average of \SI{750}{\micro\ampere} defines $I_C$. The retrapping current of approximately \SI{90}{\micro\ampere} is not used. Dividing by the nominal 200~nm by 1000~nm cross section gives $J_C \approx \SI{0.375}{\mega\ampere\per\centi\meter\squared}$. For comparison, the design value of $J_C$ for unimplanted Nb in the MIT Lincoln Laboratory process is approximately 30~MA/cm$^2$~\cite{tolpygo2014,Tolpygo2015}, so the implanted line carries about 80 times less current density than the film. The hysteresis between switching and retrapping is consistent with self heating of the implanted region.

Figure~\ref{fig:IcT}(b) shows $J_C$ of the Experiment 1 lines as a function of temperature for both electrode widths at each fluence. $J_C$ decreases with increasing temperature and approaches zero near $T_C$ of the line, and increasing the fluence moves the curves to lower temperature. At $5\times10^{12}$~ions/cm, the 750~nm electrode has $J_C = 0.08$~MA/cm$^2$ at 4.2~K, interpolated between the measurements at 3.9 and 4.3~K, and at $1\times10^{13}$~ions/cm the lines are resistive at 4.2~K. At this highest fluence, $J_C$ at 2~K is approximately two orders of magnitude below that of the unimplanted film, although $T_C$ is reduced by only about 60\%. A reduction of $J_C$ much larger than the reduction of $T_C$ was also observed in ion irradiated YBCO~\cite{White1988}.

Figure~\ref{fig:IcT}(c) shows $J_C$ of the lines in the 100~nm film of Experiment 2 for electrodes 500 to 1000~nm wide. As in Experiment 1, $J_C$ decreases toward zero near $T_C$ of the line, and increasing the fluence moves the curves to lower temperature. The curves at $5\times10^{11}$~ions/cm end near 5.5~K, below which the electrode switched before the line and $J_C$ of the line was not determined. At $1\times10^{12}$~ions/cm, $J_C$ at 4.2~K is 0.25, 0.32, and 0.33~MA/cm$^2$ on three electrodes on both sides of the chip. The spread of about 30\% between electrodes sets the margin used in Section~\ref{sec:moats}. At $5\times10^{12}$~ions/cm, the lines carry no supercurrent at 4.2~K. No intermediate fluence was written, so the fluence at which the supercurrent vanishes at 4.2~K lies between $1\times10^{12}$ and $5\times10^{12}$~ions/cm.

\subsection{TDGL Simulation}\label{sec:tdgl}

\begin{figure}[!tbp]
    \centering
    \includegraphics[width=.8\columnwidth]{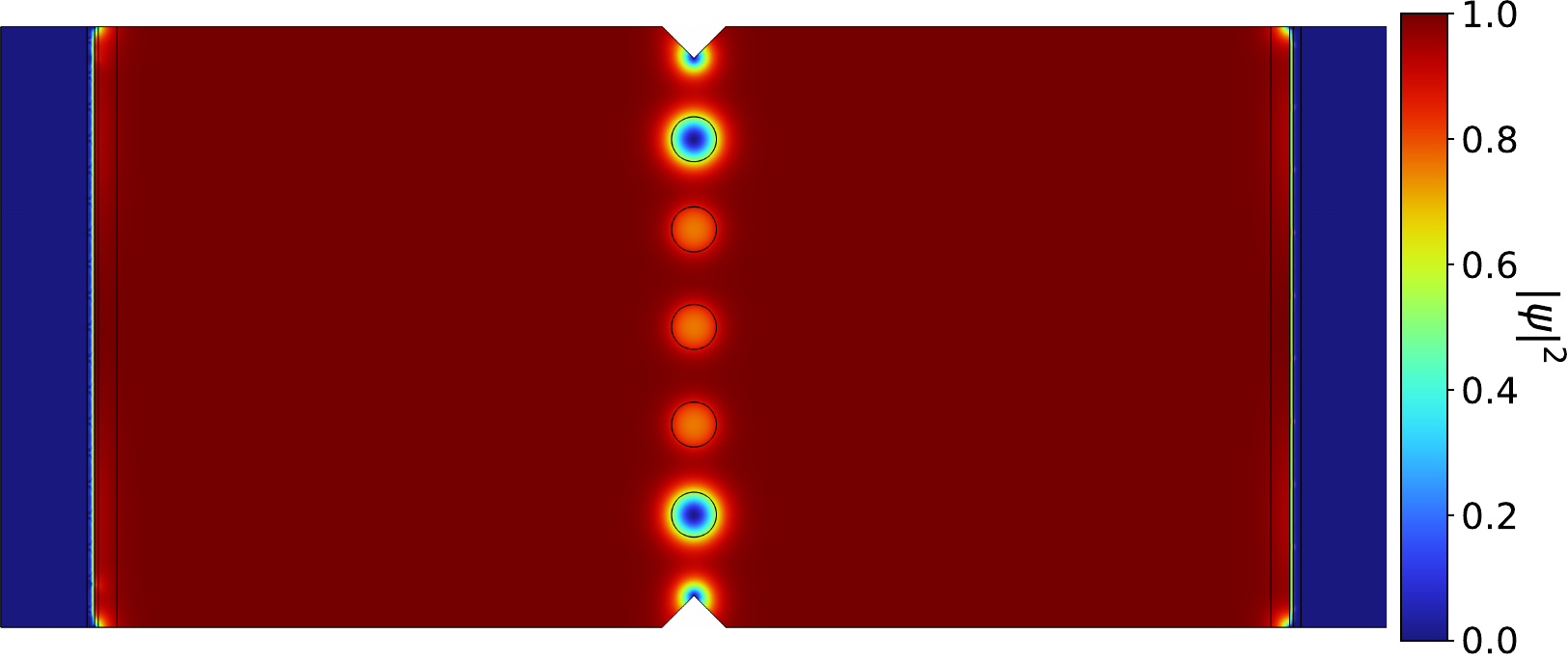}
    \caption{Two dimensional time dependent Ginzburg-Landau (TDGL) simulation of vortex trapping in a superconducting Nb strip containing circular silicon implanted moats arranged along its center. The color scale represents the squared magnitude of the superconducting order parameter, $|\Psi|^2$. Normal metal contact regions are located at the left and right ends of the strip, and notches along the upper and lower boundaries facilitate vortex nucleation. The simulation shows vortex capture and confinement within an implanted moat after $2000\tau_0$.}
    \label{fig:TDGL}
\end{figure}

To investigate vortex trapping in implanted Nb, we performed two dimensional time dependent Ginzburg-Landau (TDGL) simulations of a superconducting strip containing circular implanted moats along its center (Fig.~\ref{fig:TDGL}). We simulated the lowest fluence that produced a separate line transition, $5\times10^{11}$~ions/cm in Experiment 2, or $4.5\times10^{17}$~ions/cm$^2$ in the beam spot. This fluence reduced $T_C$ of the line from approximately 9~K to 7.4~K (Section~\ref{sec:rt}) and sputtered a trench only about 8~nm deep (Fig.~\ref{fig:trench}), so the implanted region is modeled as a weakened superconductor of the full film thickness rather than as a hole. The simulations were performed at 4.2~K. The local strength of superconductivity enters the equations through the coefficient $\epsilon = (T_C^{\prime} - T)/(T_C - T)$, where $T_C^{\prime}$ is the local transition temperature, so $\epsilon = 1$ in the unimplanted Nb and $\epsilon = (7.4 - 4.2)/(9 - 4.2) = 0.667$ in the implanted regions. Time is measured in units of the reference time $\tau_0 = 0.1005$~ps, and $\eta$ is the order parameter relaxation time in units of $\tau_0$. Because the relaxation time scales as $1/(T_C^{\prime} - T)$, $\eta = 6.22$ in the Nb and $\eta = 9.33$ in the implanted regions. Small notches along the strip boundaries facilitate vortex nucleation.

The vortex nucleated at a notch and propagated through the strip toward the implanted regions. Within a simulation time of $2000\tau_0$, it was captured and remained confined within an implanted moat. The simulation indicates that a region of locally suppressed superconductivity can capture and hold a vortex. The model is two dimensional, and the pinning strength remains to be measured.

\subsection{Implanted Moats}\label{sec:moats}

\begin{figure}[!tbp]
 \centering
 \includegraphics[width=.9\columnwidth]{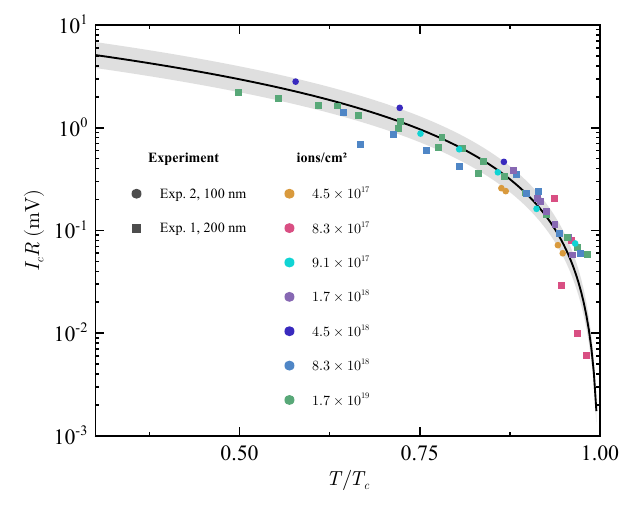}
 \caption{Product of the critical current and the resistance of the implanted line, $I_CR$, as a function of the temperature divided by the transition temperature of the line. Circles are the 100~nm film of Experiment 2, and squares are the 200~nm film of Experiment 1. The fit uses the 100 and 200~nm films of Experiment 2, and the 200~nm film of Experiment 2 is omitted from the plot for clarity. Colors give the area fluence in the beam spot in ions/cm$^2$ with the same scale as Fig.~\ref{fig:rt}. The solid line is $I_CR = V_0(1-T/T_C)^n$ fitted to the Experiment 2 points, with $V_0 = 6.0$~mV and $n = 1.39$. The gray band is the scatter of that fit, a factor of 1.9.}
 \label{fig:collapse}
\end{figure}

A designer of implanted moats needs $I_C$ of the implanted region at the operating temperature without measuring $I$-$V$ characteristics for every fluence and geometry. Figure~\ref{fig:collapse} shows the product $I_CR$ of the implanted line as a function of $T/T_C$, where $T_C$ is that of the line. For Experiment 2, $I_C$ is the first switching current, $R$ is the line resistance at 8.5~K, and $T_C$ is the midpoint of the $I$-$V$ bracket of Section~\ref{sec:meas}. For Experiment 1, $R$ is the high bias slope of each $I$-$V$ characteristic and $T_C$ is the temperature at which a linear fit of $J_C(T)$ extrapolates to zero. The 58 points of the 100 and 200~nm films of Experiment 2 follow
\begin{equation}
I_CR = V_0\left(1-\frac{T}{T_C}\right)^n,
\label{eq:collapse}
\end{equation}
where $V_0$ is the product extrapolated to zero temperature and $n$ is the exponent. The fit gives $V_0 = 6.0$~mV and $n = 1.39$, and the points scatter about the fit by a factor of 1.9, which we define as the standard deviation of the logarithm of the ratio of each point to the fit. The Experiment 1 points average 0.75 times the fit, with 84\% within the same scatter. This is the useful result of Fig.~\ref{fig:collapse}. One $R(T)$ sweep of an implanted region gives $T_C$ and $R$, and (\ref{eq:collapse}) then gives $I_C$ at any temperature to within a factor of 1.9, for both write modes, the 100 and 200~nm films, and two orders of magnitude in fluence.

The collapse is unexpected at first sight, because the lines differ in fluence, trench depth, and film thickness. We attribute it to two properties of the product. First, the geometry cancels. $I_C = J_C A_l$ and $R = \rho_{\mathrm{eff}}\, l/A_l$, where $A_l$ is the cross section of the line, so $I_CR = J_C\rho_{\mathrm{eff}}\, l$ depends neither on the electrode width nor on the thickness that remains under the trench, and $l$ is set by the straggle at 70~keV for every line. Second, the dependence on the material is weak. In the dirty limit, the depairing current density scales as $T_C^{3/2}\rho^{-1/2}$~\cite{Tinkham1996}, so $I_CR \propto l\,T_C^{3/2}\rho^{1/2}$ and the decrease of $T_C$ with fluence is largely compensated by the increase of $\rho$. With the $T_C$ and $\rho_{\mathrm{eff}}$ of the Experiment 1 lines, $T_C^{3/2}\rho_{\mathrm{eff}}^{1/2}$ varies by a factor of 1.8 between the lowest and highest fluence, within the scatter of the fit. The exponent of 1.39 is close to the Ginzburg-Landau depairing value of 3/2 near $T_C$, which supports this picture. The 0.75 offset of Experiment 1 is not understood and merits additional study.

A pinning site does not need to be normal, because any local reduction of $J_C$ lowers the energy of a vortex at that site. In a 100~nm film at 70~keV, the lowest fluence that produced a separate line transition, $4.5\times10^{17}$~ions/cm$^2$ in the beam spot, reduced $T_C$ of the implanted region to 7.4~K and removed only about 8~nm of the film. At $9.1\times10^{17}$~ions/cm$^2$, $J_C$ at 4.2~K is 0.25 to 0.33~MA/cm$^2$, about 1\% of the unimplanted film, and at $4.5\times10^{18}$~ions/cm$^2$ the region carries no supercurrent at 4.2~K but the beam has sputtered a trench about 61~nm deep. In a 200~nm film at 70~keV the ions stop in the upper half of the film, and the only fluence that modified the full thickness, $1.7\times10^{19}$~ions/cm$^2$ in a single pass, sputtered 87~nm of the film. A 200~nm film therefore needs a higher energy, approximately 162~keV for a range of 100~nm.

The TDGL simulation of Section~\ref{sec:tdgl} shows that the region written at the lowest fluence captures and holds a vortex. We therefore suggest, for implanted moats in a 100~nm Nb ground plane at 4.2~K, 70~keV Si at an area fluence of about $5\times10^{17}$~ions/cm$^2$. This is the lightest implant that produced a separate transition, it leaves the film nearly intact, and the simulation indicates that it is sufficient to pin a vortex. Higher fluences suppress $J_C$ further but sputter the film, and a moat does not need to be dug through it.

A broad beam sets the area density of implanted ions directly, which for a focused line is the line fluence divided by the 53~nm lateral extent of Section~\ref{sec:3b}. In this measure the suggested implant, $5\times10^{11}$~ions/cm, is approximately $1\times10^{17}$~ions/cm$^2$. For comparison, nitrogen implantation of 120~nm thick Nb films at 100~keV reduces $T_C$ from 9~K to a minimum of 3~K at 1.0 to $1.5\times10^{17}$~ions/cm$^2$~\cite{Johnson1990}, and SRIM gives 2.2 times as many vacancies per ion for 70~keV Si as for 100~keV N, so the two results are of the same order. A focused line is flanked by unimplanted Nb, so this fluence is an upper bound for a broad beam. The conversion has not been tested, and a broad beam implant must confirm it.

\section{Conclusion}

In conclusion, we have written silicon implanted lines in niobium films 50, 100, and 200~nm thick with a focused ion beam in two write modes and measured how the transition temperature, resistance, and critical current density of the lines depend on the fluence. The transition temperature of the line falls and its resistance rises with fluence while the surrounding film stays near 9~K. The critical current density falls by about two orders of magnitude before the line becomes resistive at 4.2~K. The beam sputters a trench whose depth is set by the area fluence in the beam spot. The product $I_CR$ of the lines from both experiments follows one power law in $T/T_C$ to within a factor of 1.9, so the critical current of a line at any temperature can be estimated from its $T_C$ and resistance. A time dependent Ginzburg-Landau simulation indicates that an implanted region can capture and hold a vortex. For implanted moats in a 100~nm Nb ground plane at 4.2~K, we suggest 70~keV Si at an area fluence of about $5\times10^{17}$~ions/cm$^2$ in the beam spot, the lightest implant that produced a separate transition, with an estimated broad beam equivalent of the order of $10^{17}$~ions/cm$^2$. A 200~nm film requires a higher ion energy, because the only fluence that modified its full thickness at 70~keV also sputtered almost half the film. The conversion from line fluence to broad beam fluence is an estimate, and flux trapping was not measured. Broad beam implants and magnetic imaging of implanted structures will determine the fluence and geometry that best pin flux in niobium digital circuits.

\section*{Acknowledgment}

The authors thank Peter Gnauck, Andreas Remscheid, Torsten Richter, and Mahesh Hingorani of Raith GmbH for supporting logistics in both experiments with the VELION system. The authors also thank MIT Lincoln Laboratory for fabricating the single layer niobium samples and the Department of Electrical and Computer Engineering at the University of California, Riverside.  This work was supported by the Army Research Office (ARO) under Grant W911NF-24-1-0153 and by the Department of Energy National Nuclear Security Administration (NNSA) under Grant DE-NA0004106.

\bibliographystyle{unsrt}
\bibliography{biblio}

\begin{thebibliography}{10}

\bibitem{Shehabi2024}
Arman Shehabi, Sarah~Josephine Smith, Alex Hubbard, Alexander Newkirk, Nuoa
  Lei, Md~AbuBakar Siddik, Billie Holecek, Jonathan~G. Koomey, Eric~R. Masanet,
  and Dale~A. Sartor.
\newblock 2024 {United States} data center energy usage report.
\newblock Technical Report LBNL-2001637, Lawrence Berkeley National Laboratory,
  Berkeley, CA, USA, 2024.

\bibitem{Holmes2013}
D.~Scott Holmes, Andrew~L. Ripple, and Marc~A. Manheimer.
\newblock Energy-efficient superconducting computing: Power budgets and
  requirements.
\newblock {\em IEEE Transactions on Applied Superconductivity}, 23(3):1701610,
  2013.

\bibitem{Tolpygo2016}
Sergey~K. Tolpygo.
\newblock Superconductor digital electronics: Scalability and energy efficiency
  issues.
\newblock {\em Low Temperature Physics}, 42(5):361--379, 2016.

\bibitem{Scalapino1967}
C.~S. Owen and D.~J. Scalapino.
\newblock Vortex structure and critical currents in {Josephson} junctions.
\newblock {\em Phys. Rev.}, 164:538--544, Dec 1967.

\bibitem{golod2015}
Taras Golod, Adriana Iovan, and Vladimir~M. Krasnov.
\newblock Single {Abrikosov} vortices as quantized information bits.
\newblock {\em Nature Communications}, 6:8628, 2015.

\bibitem{Jeffery1995}
Mark Jeffery, T.~Van~Duzer, J.~R. Kirtley, and M.~B. Ketchen.
\newblock Magnetic imaging of moat-guarded superconducting electronic circuits.
\newblock {\em Applied Physics Letters}, 67(12):1769--1771, 09 1995.

\bibitem{Semenov2016}
Vasili~K. Semenov and Mikhail~M. Khapaev.
\newblock How moats protect superconductor films from flux trapping.
\newblock {\em IEEE Transactions on Applied Superconductivity}, 26(3):1--10,
  2016.

\bibitem{Kapur2026Mitigation}
Rohan~T. Kapur, Sergey~K. Tolpygo, Alex Wynn, Pauli Kehayias, Adam~A. Libson,
  Collin~N. Muniz, Michael~J. Gold, Justin~L. Mallek, Danielle~A. Braje, and
  Jennifer~M. Schloss.
\newblock Mitigation of magnetic flux trapping in superconducting electronics
  using moats.
\newblock {\em Phys. Rev. Appl.}, 26:034038, Sep 2026.

\bibitem{Kapur2026Flux}
Rohan~T. Kapur, Pauli Kehayias, Sergey~K. Tolpygo, Adam~A. Libson, George
  Haldeman, Collin~N. Muniz, Alex Wynn, Nathaniel~J. O'Connor, Neel~A. Parmar,
  Ryan Johnson, Andrew~C. Maccabe, John Cummings, Justin~L. Mallek, Danielle~A.
  Braje, and Jennifer~M. Schloss.
\newblock Flux-trapping characterization for superconducting electronics using
  a cryogenic widefield {N-V} diamond microscope.
\newblock {\em Phys. Rev. Appl.}, 25:044073, Apr 2026.

\bibitem{Yamae2025}
Taiki Yamae, Yuki Hironaka, Shuichi Nagasawa, Yuki Yamanashi, Nobuyuki
  Yoshikawa, and Naoki Takeuchi.
\newblock Rapid single-flux-quantum and adiabatic quantum-flux-parametron cell
  libraries using a 1 {kA/cm$^2$} niobium fabrication process.
\newblock {\em Scientific Reports}, 15(1):41429, 2025.

\bibitem{White1988}
Alice~E. White, K.~T. Short, R.~C. Dynes, A.~F.~J. Levi, M.~Anzlowar, K.~W.
  Baldwin, P.~A. Polakos, T.~A. Fulton, and L.~N. Dunkleberger.
\newblock Controllable reduction of critical currents in
  {YBa$_2$Cu$_3$O$_{7-\delta}$} films.
\newblock {\em Applied Physics Letters}, 53(11):1010--1012, 1988.

\bibitem{cybart2015}
Shane~A. Cybart, E.~Y. Cho, T.~J. Wong, Bj{\"o}rn~H. Wehlin, Meng~K. Ma, Chuong
  Huynh, and R.~C. Dynes.
\newblock Nano {Josephson} superconducting tunnel junctions in
  {YBa$_2$Cu$_3$O$_{7-\delta}$} directly patterned with a focused helium ion
  beam.
\newblock {\em Nature Nanotechnology}, 10(7):598--602, 2015.

\bibitem{Cybart2014}
Shane~A. Cybart, Patricia X.~T. Yen, Ethan~Y. Cho, Jeong~Uk Huh, V.~N.
  Glyantsev, Christopher~S. Yung, Brian Moeckly, Jeffrey~W. Beeman, and
  Robert~C. Dynes.
\newblock Comparison of {Y-Ba-Cu-O} films irradiated with helium and neon ions
  for the fabrication of {Josephson} devices.
\newblock {\em IEEE Transactions on Applied Superconductivity}, 24(4):1--5,
  2014.

\bibitem{Ogawa2021}
Shinichi Ogawa, Gregor Hlawacek, Fayong Liu, Shu Nakaharai, Nirjhar Sarkar, and
  Shane Cybart.
\newblock Helium-ion microscopy and its applications in nanoscale devices.
\newblock In {\em Outlooking beyond Nanoelectronics and Nanosystems}, pages
  139--213. Jenny Stanford Publishing, 2024.

\bibitem{Brooks2025}
Alexander~J. Brooks, Jay~C. LeFebvre, Joseph~A. Forman, Sreekar Vattipalli,
  Alex Belianinov, Michael Titze, and Shane~A. Cybart.
\newblock Electrical characterization of {Co} implanted {Y-Ba-Cu-O} thin films.
\newblock {\em IEEE Transactions on Applied Superconductivity}, 35(5):1--3,
  2025.

\bibitem{Hoflich2023}
Katja H{\"o}flich, Gerhard Hobler, Francis~I. Allen, Tonio Wirtz, Gemma Rius,
  Lisa McElwee-White, Arkady~V. Krasheninnikov, Michael Schmidt, Ivo Utke,
  Nikolaus Klingner, et~al.
\newblock Roadmap for focused ion beam technologies.
\newblock {\em Applied Physics Reviews}, 10(4):041311, 2023.

\bibitem{Williams1998}
J.~S. Williams.
\newblock Ion implantation of semiconductors.
\newblock {\em Materials Science and Engineering: A}, 253(1--2):8--15,
  September 1998.

\bibitem{Tolpygo2015}
Sergey~K. Tolpygo, Vladimir Bolkhovsky, Terence~J. Weir, Leonard~M. Johnson,
  Mark~A. Gouker, and William~D. Oliver.
\newblock Fabrication process and properties of fully-planarized deep-submicron
  {Nb/Al-AlO$_x$/Nb} {Josephson} junctions for {VLSI} circuits.
\newblock {\em IEEE Transactions on Applied Superconductivity}, 25(3):1--12,
  2015.

\bibitem{Datesman2005}
Aaron~M. Datesman, Jonathan~C. Schultz, Thomas~W. Cecil, Christine~M. Lyons,
  and Arthur~W. Lichtenberger.
\newblock Gallium ion implantation into niobium thin films using a focused-ion
  beam.
\newblock {\em IEEE Transactions on Applied Superconductivity},
  15(2):3524--3527, 2005.

\bibitem{Bischoff2005}
L.~Bischoff.
\newblock Alloy liquid metal ion sources and their application in mass
  separated focused ion beams.
\newblock {\em Ultramicroscopy}, 103(1):59--66, 2005.
\newblock Proceedings of the Ninth Conference on Frontiers of Electron
  Microscopy in Materials Science.

\bibitem{Bischoff2016}
L.~Bischoff, P.~Mazarov, L.~Bruchhaus, and J.~Gierak.
\newblock Liquid metal alloy ion sources---an alternative for focussed ion beam
  technology.
\newblock {\em Applied Physics Reviews}, 3(2):021101, 2016.

\bibitem{Ziegler2010}
James~F. Ziegler, M.D. Ziegler, and J.P. Biersack.
\newblock {SRIM – The stopping and range of ions in matter (2010)}.
\newblock {\em Nucl. Instrum. Methods Phys. Res. Sec. B},
  268(11--12):1818--1823, 2010.

\bibitem{Tzanev1992}
S.~J. Tzanev, D.~S. Karpuzov, and A.~A. Kosyachkov.
\newblock Influence of ion channeling on backscattering and sputtering yields
  of niobium single crystals.
\newblock {\em Nuclear Instruments and Methods in Physics Research Section B:
  Beam Interactions with Materials and Atoms}, 71(4):381--386, 1992.

\bibitem{Moller2014}
Wolfhard M{\"o}ller.
\newblock {TRI3DYN} -- collisional computer simulation of the dynamic evolution
  of 3-dimensional nanostructures under ion irradiation.
\newblock {\em Nuclear Instruments and Methods in Physics Research Section B:
  Beam Interactions with Materials and Atoms}, 322:23--33, 2014.

\bibitem{Kohopaa2024}
Katja Kohopää, Alberto Ronzani, Robab~Najafi Jabdaraghi, Arijit Bera, Mário
  Ribeiro, Dibyendu Hazra, Jorden Senior, Mika Prunnila, Joonas Govenius,
  Janne~S. Lehtinen, and Antti Kemppinen.
\newblock Effect of ion irradiation on superconducting thin films.
\newblock {\em APL Materials}, 12(7):071101, 07 2024.

\bibitem{Bishop1985}
D.~J. Bishop, E.~G. Spencer, and R.~C. Dynes.
\newblock The metal-insulator transition in amorphous {Nb:Si}.
\newblock {\em Solid-State Electronics}, 28(1--2):73--79, 1985.

\bibitem{tolpygo2014}
Sergey~K Tolpygo, V~Bolkhovsky, T~Weir, L~M Johnson, W~D Oliver, and M~A
  Gouker.
\newblock Deep sub-micron stud-via technology of superconductor vlsi circuits.
\newblock {\em Superconductor Science and Technology}, 27(2):025016, jan 2014.

\bibitem{Tinkham1996}
Michael Tinkham.
\newblock {\em Introduction to Superconductivity}.
\newblock McGraw-Hill, New York, 2nd edition, 1996.

\bibitem{Johnson1990}
J.~P. Johnson and W.~G. Moulton.
\newblock Superconducting and normal properties of {NbN} produced by ion
  implantation of niobium thin films.
\newblock {\em Radiation Effects and Defects in Solids}, 113(4):283--296, 1990.

\end{thebibliography}

\end{document}